\documentclass[pdflatex,sn-mathphys-num,iicol]{sn-jnl}

\usepackage{graphicx}%
\usepackage{xcolor}%
\usepackage{multirow}%
\usepackage{amsmath,amssymb,amsthm,bm}
\usepackage{mathtools,physics}%
\usepackage[title]{appendix}%
\usepackage{xcolor}%
\usepackage{dsfont}%
\usepackage{textcomp}%
\usepackage{manyfoot}%
\usepackage{booktabs}%
\usepackage{algorithm}%
\usepackage{algorithmicx}%
\usepackage{algpseudocode}%
\usepackage{listings}%
\usepackage{hyperref}%
\usepackage{booktabs}%
\usepackage{siunitx}%
\usepackage{ulem}%
\usepackage[switch]{lineno}

\theoremstyle{thmstyleone}%
\newtheorem{theorem}{Theorem}%
\theoremstyle{thmstyletwo}%
\theoremstyle{thmstylethree}%
\newcommand{\g}{\mathcal G}%
\newcommand{\z}{\mathcal Z}%
\newcommand{\zkg}{Z^\mathrm{KG}}
\newcommand{\zpe}{Z^\mathrm{PE}}
\newcommand{\xkg}{X^\mathrm{KG}}
\newcommand{\xpe}{X^\mathrm{PE}}
\newcommand{\pkey}{p^\mathrm{K}}%
\newcommand{\sdelta}{\Delta_\mathrm{S}}%
\newcommand{\Nc}{N_\mathrm{c}}%
\newcommand{\etaD}{\eta_\mathrm{D}}%
\newcommand{\etaCh}{\eta_\mathrm{Ch}}%
\newcommand{\lambdaEC}{\lambda_\mathrm{EC}}%
\newcommand{\nuel}{\nu_\mathrm{el}}%
\newcommand{\mc}[1]{\mathcal{#1}}%
\newcommand{\ePA}{\varepsilon_\mathrm{PA}}%
\newcommand{\eEC}{\varepsilon_\mathrm{EC}}%
\newcommand{\config}{\boldsymbol{\varphi}}%

\def\equationautorefname~#1\null{Eq.~(#1)\null}%
\newlength{\singlecolumnfigurewidth}%
\newlength{\doublecolumnfigurewidth}%
\begin{document}


\title{Experimental demonstration of finite-size general security via discrete-modulated CVQKD with real time postprocessing}


\author*[1]{\fnm{Sven} \sur{Bodenstedt}}
\email{sven.bodenstedt@luxquanta.com}
\equalcont{These authors contributed equally to this work.}

\author*[1]{\fnm{Carlos} \sur{Pascual-Garc\'ia}}
\email{carlos.pascual@luxquanta.com}
\equalcont{These authors contributed equally to this work.}

\author[1]{\fnm{Nil} \sur{Canta i Pujol}}
\author[1]{\fnm{Mart\'i} \sur{Sales-Moragues}}
\author[1,2]{\fnm{Mariana} \sur{Navarro}}
\author[1]{\fnm{Pau} \sur{G\'omez Kabelka}}
\author[1]{\fnm{Sebastián} \sur{Etcheverry}}
\author[1]{and \fnm{Saeed} \sur{Ghasemi}}

\affil[1]{\orgname{Luxquanta Technologies S.L.}, \orgaddress{\street{Av. Joan Carles I, 30, 1º1ª}, \city{L’Hospitalet de Llobregat}, \postcode{08908}, \state{Barcelona}, \country{Spain}}}
\affil[2]{\orgdiv{ICFO - Institut de Ciences Fotoniques}, \orgname{The Barcelona Institue of Science and Technology}, \orgaddress{\street{Av. Carl Friedrich Gauss, 3}, \city{Castelldefels}, \postcode{08860}, \state{Barcelona}, \country{Spain}}}

\abstract{%
Continuous-variable quantum key distribution (CVQKD) is compatible with telecommunication infrastructure, but implementing composable security with experimentally practical resources has remained challenging, particularly for discrete-modulated (DM) protocols.
We report the first experimental demonstration of a DM CVQKD system that generates composable secret keys against general attacks with finite-size block lengths on the order of $\sim 10^{6}$ rounds via a quadrature phase shift keying (QPSK) system.
Our implementation follows a variable-length, general security framework enabled by modern entropy accumulation techniques and conic optimization, whose experimental pipeline allows real-time operation on near-commercial hardware.
}

\maketitle

Quantum key distribution (QKD) \cite{BB84,E91,renner2006security} stands as the most advanced application of quantum communication \cite{pirandola2019advances}, providing a mechanism to establish cryptographic keys whose security relies on the laws of quantum mechanics. For such keys to be of practical use, their security must be composable, i.e. preserved when the key is consumed by any subsequent cryptographic application, and it must hold for a finite numbers of quantum signals actually exchanged and against the most general attacks an eavesdropper can perform \cite{renner2006security}. To translate this theoretical security into scalable, high-rate networks, continuous-variable QKD (CVQKD) offers a highly appealing architecture due to its direct compatibility with standard telecommunications infrastructure \cite{Usenko2026Continous,zhang2024continuousvariable}.

Protocols based on Gaussian-modulated distributions of coherent states \cite{Laudenbach2018Continuous} have traditionally been the main subject of experimental study for CVQKD, with demonstrations of security against collective attacks beyond $\SI{100}{km}$ \cite{Huang2016londistanceCVQKD, Zhang2020,Hajomer2024} and, in the finite-size composable setting, over $\SI{20}{km}$ \cite{Jain2022practicalCVQKD}. To date, however, all CVQKD implementations with coherent states have been limited to security against collective attacks, in which the eavesdropper is assumed to interact identically and independently with each transmitted state. For Gaussian-modulated protocols, security against general attacks can in principle be obtained through a reduction to collective attacks, but this reduction degrades the security parameter with respect to the block size and requires a computationally demanding symmetrization of the data \cite{Jain2022practicalCVQKD}, which has so far prevented its experimental demonstration.
 
On the other hand, discrete-modulated (DM) CVQKD  \cite{leverrier2011continuous,ghorai2019asymptotic,kaur2019asymptotic} has attracted considerable interest because it significantly reduces computational and theoretical overhead compared with Gaussian-modulated protocols. By encoding information into a finite constellation of coherent states and discretizing the measurement outcomes, DM CVQKD overcomes the heavy digital signal processing and complex reconciliation inherent to infinite-dimensional registers \cite{Leverrier2009Unoconditional} and, more importantly, makes the protocol compatible with entropy accumulation techniques \cite{arqand2025MEAT, tupkary2026rigorouscompletesecurityproof}, which can provide finite-size security against general attacks directly without reduction to collective attacks. 

The practical viability of DM CVQKD, particularly beyond metropolitan scales, has been questioned for years \cite{leverrier2023information} as this approach is constrained by the efficiency of information reconciliation  and the experimental implementability of finite-size security methods. 
Recent experiments have reported finite-size composable security against collective attacks for both Gaussian-modulated \cite{Huang2016londistanceCVQKD,Jain2022practicalCVQKD} and DM protocols \cite{Hajomer2025,Mingze2026HighRate}. Extending these works to security against general attacks while retaining practical block sizes and efficient classical postprocessing remains the outstanding challenge for CVQKD with coherent states.

In this work, we report the first experimental demonstration of CVQKD using coherent states producing finite-size composable secret keys secure against general attacks in real time via DM CVQKD. Based on the standard quadrature phase shift keying (QPSK) modulation, our demonstration crucially introduces a variable-length decision (VLD)  \cite{Tupkary2024, arqand2025MEAT}. This allows the protocol to dynamically adjust the final secret key length based on the statistical observations of the shared classical information, together with a database of verifiable precomputed key lengths for a sharp, real time estimation. 

By further combining advanced security proofs based on the marginal-constrained entropy accumulation theorem (MEAT) \cite{arqand2025MEAT, tupkary2026rigorouscompletesecurityproof}, with efficient conic optimization methods \cite{Navarro2025,NavarroPostsel2026}, we successfully extract secret keys from comparatively small data blocks ($\sim 10^6$) for distances up to $\SI{10}{km}$. For larger block sizes ($\sim 10^8$) we achieved continuous secret key rates from $\sim \SI{10}{kbits/s}$ at \SI{5}{km} to $\sim \SI{0.1}{kbits/s}$ at \SI{40}{km} on average. These results show that composable security against general attacks can be achieved in a practical CVQKD system with coherent states, resolving a long-standing open problem of the field.

\section*{Results} \label{Sec:Results}

\subsection*{Protocol description} \label{Sec:Protocol}

\begin{figure*}[h]
    \centering
    \includegraphics[width=\textwidth]{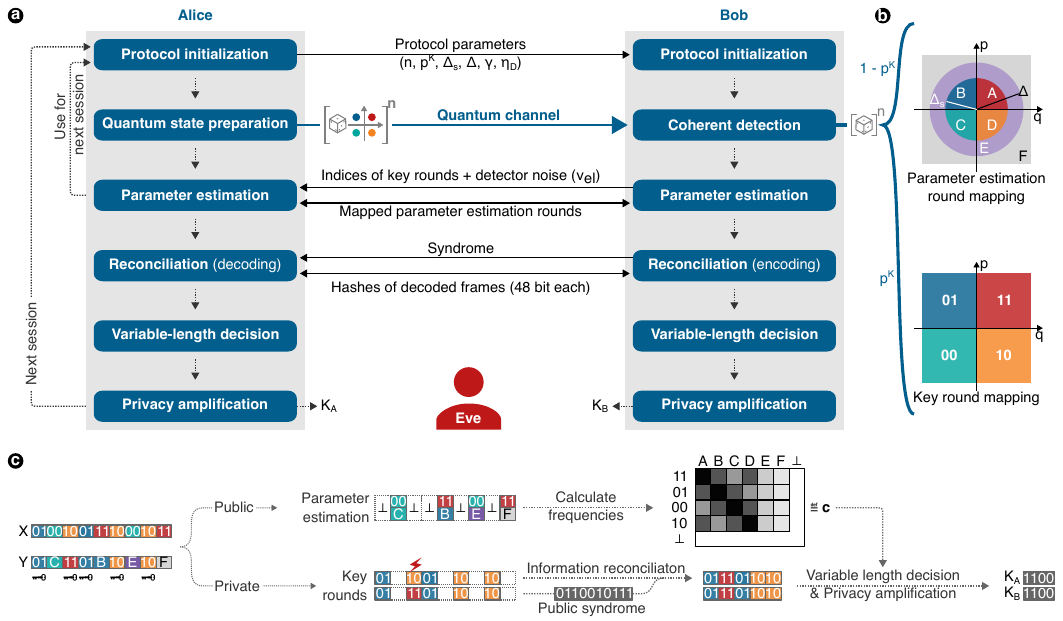}
    \caption{
    Implementation overview.
    (a) Main steps of the protocol: initialization, quantum state preparation and coherent detection, parameter estimation, reconciliation, VLD, and privacy amplification. Black arrows between Alice and Bob represent an authenticated classical channel, whereas the blue arrow represents the quantum channel.
    (b) In each round, Alice randomly prepares one of the four coherent states of the QPSK alphabet \{\texttt{00}, \texttt{01}, \texttt{10}, \texttt{11}\}. Bob then decides, with probability $\pkey$, whether to treat the round as a key round. If so, he maps his heterodyne measurement outcome to one of the four QPSK letters according to the quadrant that it occupies in phase space. Otherwise, the round is used for parameter estimation, and the outcome is mapped to one of six regions (\texttt{A}, \texttt{B}, \texttt{C}, \texttt{D}, \texttt{E}, \texttt{F}) defined by the protocol parameters $\sdelta$ and $\Delta$.
    (c) Once all rounds have been measured, Alice holds the array $X$ of prepared symbols, while Bob holds the array $Y$ of mapped measurement outcomes together with the key-round identifiers (indicated by the little key symbols below $Y$). The thunderbolt symbol marks disagreement arising either from noise in the measurement process or from Eve's intervention. The frequency vector $\mathbf{c}$ is illustrated as a matrix, with different shades corresponding to different frequencies of observing the specific letter in the alphabet $\mathcal{C}$. It has 25 elements: $4 \times 6$ for the possible combinations during a parameter estimation round and one additional L-shaped element representing the key rounds. 
    }
    \label{fig:protocol}
\end{figure*}

The implemented protocol follows the description of Pascual-Garc\'ia \textit{et al.} \cite{PascualGarcia2025} with the security framework provided in Navarro \textit{et al.} \cite{Navarro2025}, together with a trusted noise model for the detectors from Lin \textit{et al.} \cite{LinTrusted2020}. A schematic overview is shown in \autoref{fig:protocol}. The protocol distributes QPSK-modulated coherent states, which are exchanged between Alice and Bob over a quantum channel, while all classical postprocessing is coordinated entirely via an authenticated classical channel. Every QKD session consists of the following steps:

\begin{enumerate}
    \item \textbf{Protocol initialisation.} 
    Alice and Bob agree on the parameters that characterise the QKD session. They exchange the chosen protocol parameters: total symbol count $n$, categorisation parameters $\sdelta$ and $\Delta$, key-round fraction $\pkey$; and secrecy $\varepsilon_\mathrm{PA}$ and correctness $\varepsilon_\mathrm{EC}$ parameters, which define an $(\varepsilon_\mathrm{PA} + \varepsilon_\mathrm{EC})-$secure implementation.  They also fix the calibrated experimental parameters, namely the coherent-state amplitude $\gamma$ and Bob's detector efficiency $\etaD$, which is measured beforehand and cannot be chosen freely.
    
    \item \textbf{Quantum communication.}
    In each round $j \in \{1,...,n\}$, Alice draws two bits $x \in \{\mathtt{00}, \mathtt{01},\mathtt{10},\mathtt{11}\}$ from a quantum random number generator and prepares the corresponding coherent state  $\ket{\gamma_x}$. She sends the generated state to Bob and stores her choice in $X_j$. Bob performs heterodyne detection and assigns the round to key generation with probability $\pkey$. Otherwise, the round is used for parameter estimation. He maps the outcome to four regions (key round) or six regions (parameter-estimation rounds), both defined by the amplitudes $\sdelta$ and $\Delta$ (see \autoref{fig:protocol} b), and records the result in register $Y_j$.
    
    \item \textbf{Public announcements.}
    Bob publicly announces his round designations, his outcomes $Y_j$ for parameter-estimation rounds, and the calibrated detector noise $\nuel$; Alice announces her symbols $X_j$ for all parameter-estimation rounds (\autoref{fig:protocol}\,c), such that they together build a public register $C_j$. Bob builds the raw key register using the undisclosed rounds.
    
    \item \textbf{Information reconciliation.} Alice and Bob remove the discrepancies between their keys using a reverse reconciliation LDPC scheme, such that Alice generates a guess of Bob's key. The total error correction leakage 
    is subtracted in the VLD, and each frame is validated according to a universal$_2$ hash of length $b_\mathrm{EC}$.

    \item \textbf{Variable-length decision.} Provided the publicly shared information and the reconciliation leakage, Alice and Bob evaluate the final secret key length for each valid configuration in a precomputed database of key lengths (defined in \autoref{sec:database}), and select the configuration giving the largest key length.

    \item \textbf{Privacy amplification.} Using either a pre-shared seed or exchanging one via the authenticated channel,
    Alice and Bob apply Toeplitz hashing \cite{Krawczyk1994} to compress the reconciled string to the secret-key length from the previous step, yielding an information-theoretically secure key. 
\end{enumerate}

Following protocol initialization, all classical communication and post-processing is carried out only after every quantum round has been measured, allowing a simplified security analysis according to Ref. \cite[Corollary 4.2]{arqand2025MEAT}.

To estimate the secret key length, we collect the protocol
and experimental parameters into the configuration vector
\begin{equation}
\begin{split}
    \config := & [\,
    \underbrace{\ePA, \eEC, n,\,\pkey,\,\sdelta,\,\Delta,\Nc}_{\text{chosen}}
    , \\
    &
    \underbrace{\gamma,\,\etaD,\,\nuel}_{\text{calibrated}}
    ,
    \underbrace{\etaCh,\,\xi}_{\text{estimated}}
    \,].
\end{split}
\end{equation}
Here $\xi$ is the excess noise of the channel, $\etaCh$ the channel transmittance and the photon-number cutoff $\Nc$. In \autoref{tab:parameters_const} we list the subset of parameters that are constant throughout this work.

\begin{table}[h]
    \centering
    \caption{
        Constant parameters or limits. We note that the exact correctness parameter $\eEC = m  2^{-b_\mathrm{EC}} $, with number of frames $m = \lfloor (2n \pkey_\mathrm{exp}) / L_\mathrm{LDPC} \rfloor$, depends on the number of symbols $n$ and $\pkey_\mathrm{exp}$. Whereas $\Nc$ refers to the assumed maximal number of photons that compose the coherent states.
    }
    \begin{tabular}{p{3.5cm} l r}
        \toprule
         \textbf{Parameter / Limit} & \textbf{Symbol} & \textbf{Value}
         \\
         \midrule
         Correctness parameter & $\eEC$ & $\leq\num{1e-11}$
         \\
         Secrecy parameter & $\ePA$ & $\num{9e-11}$
         \\
         Cutoff number & $\Nc$ & \num{10}
         \\
         Detection efficiency & $\etaD$ & $\leq \SI{55.0}{\percent}$
         \\
         EC encoding efficiency & $\beta_\mathrm{enc}$ & $\leq \SI{97.0}{\percent}$
         \\
         EC frame hash length & $b_\mathrm{EC}$ & \SI{48}{bit}
         \\
         Code length & $L_\mathrm{LDPC}$ & $\leq 2^{23}$ 
         \\
         \bottomrule
    \end{tabular}
    \label{tab:parameters_const}
\end{table}

\subsection*{Secret key length estimation} \label{Sec:SecretKeyRateEstimation}

To quantify the secret key length achievable with our protocol, we employ a variable-length framework \cite{Tupkary2024} (see \autoref{Sec:SKREstimate} for a quantitative description of this process according to the protocol outline). This approach allows Alice and Bob to adjust the length of their final secret key according to their observations during parameter estimation and the reconciliation leakage, while retaining composable security against general (coherent) attacks \cite{kamin25MEATsecurity}, based on the $\varepsilon-$security framework \cite{renner2006security}. In particular, it can be integrated within the general security proof based on the MEAT \cite{arqand2025MEAT,kamin25MEATsecurity}.

The secret key length is governed by a pre-defined trade-off function $f_{\config}: \mathcal{C} \to \mathbb{R}$, defined over the alphabet $\mathcal{C}$ of public information $C$. This function can be arbitrarily chosen, and we typically optimize it for a configuration $\config$. Moreover, $f_{\config}$ defines a coefficient $\kappa_{\config}$ that bounds the final secret key length through the $f$-weighted R\'enyi entropy of order $\alpha$, ${H}^{\uparrow, f}_{\alpha}$ (see \cite{arqand2025MEAT}). Similarly, the Rényi parameter $\alpha$ can be optimized for each configuration $\config$ to maximize the key length, and as such we also denote the optimized order $\alpha_{\config}$. In evaluating $\kappa_{\config}$, we impose a photon-number cutoff assumption on Bob's received state \cite{Lin2019Asymptotic}. 

Provided these definitions, we evaluate the protocol performance according to the following theorem \cite{kamin25MEATsecurity}.

\begin{theorem} \label{th:MainTheorem}
    Let $\alpha \in (1,2)$, $\ePA, \eEC \in (0,1]$, and $f_{\config} : {\mathcal{C}} \to \mathbb{R}$ a tradeoff function chosen according to $\config$, denoted as a vector $\mathbf{f}_{\config} \in \mathbb{R}^{|\mathcal{C}|}$ acting on $C$. Let further $\kappa$ be the ${H}^{\uparrow, f}_{\alpha}$-normalisation constant corresponding to the set of all states that can be produced in a single use of the quantum channel at the QKD protocol. 
    Then, the protocol is $\ePA$-secret and $\eEC$-correct, hence $(\varepsilon_\mathrm{PA} + \varepsilon_\mathrm{EC})$-secure, 
    producing a secret key whose variable length $\ell_\varphi$ is, conditioned on a successful error correction validation, 
    determined from the observed finite frequencies $\mathbf{c}$ and reconciliation leakage,
    satisfying
    \begin{align} \label{eq:VarLength}
        \ell_{\config} \leq \max\left\{0,\mathbf{v}_{\config} \cdot {\mathbf{w}}\right\}
    \end{align}
    where we define the vectors
    \begin{subequations}
    \begin{align}
        \mathbf{v}_{\config}
        :=
        & \left [ \frac{\alpha_{\config}}{\alpha_{\config}-1}, \kappa_{\config}, \mathbf{f}_{\config}, 1\right],
        \\
        \begin{split}
        \mathbf{w}
        :=
        & \Bigg[ -\log \left(\frac{1}{\ePA}\right), n, n \, \mathbf{c}, 2-\lambdaEC(\mathbf{c},\mathbf{h}) 
         \Bigg],
        \end{split}
    \end{align}
    \end{subequations}
    with $\lambdaEC (\mathbf{c},\mathbf{h})$ denoting the total amount of raw key bits lost to the adversary during information reconciliation, including frame validation, according to $\mathbf{c}$ and a frame error rate determined by the public hash register $\mathbf{h}$. Frame validation ensures $\eEC$-correctness (see \autoref{Sec:Leakage} for further details).
\end{theorem}

According to the definitions stated in this theorem, only the session data ($\mathbf{c}$, $\lambdaEC(\mathbf{c},\mathbf{h})$) vary at runtime. Unlike fixed-length realizations, this approach bypasses the need for Alice and Bob to validate their statistics according to finite-size estimators, which also yields a runtime advantage, as it allows an offline pre-estimation of the possible secret keys that can be distilled.

The only configuration-dependent objects in \autoref{eq:VarLength}, namely, the trade-off vector $\mathbf{f}_{\config}$, the constant $\kappa_{\config}$, and $\alpha_{\config}$, are independent of the session data: they depend on $\config$ alone, whereas the outcomes of the quantum measurements only enter the VLD through the finite frequency distribution $\mathbf{c}$ and the reconciliation leakage $\lambdaEC(\mathbf{c},\mathbf{h})$. This separation suggests that the $\config$-dependent objects could be prepared ahead of time. However, because the configuration is dictated by the channel and cannot be freely chosen, the exact $\mathbf{v}_\mathrm{exp}$ is not available in advance. In the following, we show how a precomputed database, combined with a controlled mismatch that preserves composable security, resolves this limitation.

\subsection*{Precomputed configuration database} \label{sec:database}
\begin{figure}[h]
    \centering
    \includegraphics[width=\columnwidth]{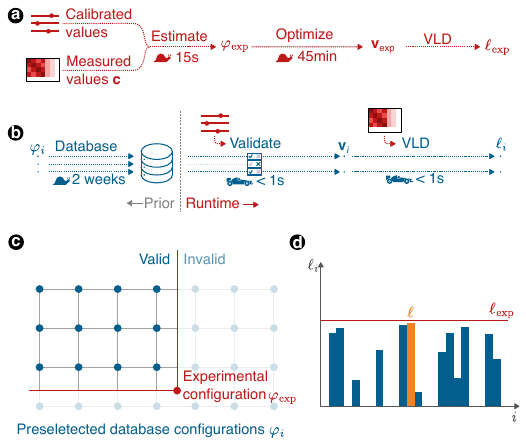}
    \caption{
        Schematic overview of the database approach for the VLD.
        (a) In a calculation at runtime, the experimentally realized configuration $\boldsymbol{\varphi}_\mathrm{exp}$ needs to be estimated from the measured data and calibrated values before the variables needed for the VLD $\mathbf{v}_\mathrm{exp}$ can be calculated. The whole process takes about \SI{45}{min} on our experimental control hardware. 
        (b) The pre-computation approach fills a database before the actual experiment. This step is computationally expensive and can take days to weeks. At runtime, the calibrated values are used to filter the pre-stored values $\mathbf{v}_i$ based on validity criteria. For each valid configuration, the VLD is performed using the measured data and the reconciliation leakage.
        (c) A multi-dimensional grid of database configurations (shown here in 2D for illustration) is precomputed. The experimental configuration $\config_\mathrm{exp}$ generally falls between grid points, and only a subset of the surrounding configurations are valid. Among these, the one yielding the highest secret key length is selected.
        (d) By design, $\ell_\mathrm{exp}$ calculated with approach (a) serves as an upper bound for the calculated key lengths $\ell_i$. Using the database approach therefore does not violate composable security. Out of all calculated key lengths, the highest (orange) is eventually selected as the secret key length.
    }
    \label{fig:database}
\end{figure}

In order to ease the notation, let us henceforth refer to specific configurations such as $\config_i$ via the shorthand $z_i := z_{\config_i}$ for any variable $z$ parameterized by $\config$ (e.g., $\mathbf{v}_{\config}$). In principle, the per-session realized configuration $\config_\mathrm{exp}$ is not known beforehand. It could be estimated after the session from the parameter-estimation frequencies $\mathbf{c}$, with the corresponding $\mathbf{v}_\mathrm{exp}$ computed at runtime. This approach is illustrated in \autoref{fig:database} (a). However, this optimization is computationally expensive and would dominate the session, thereby reducing the achievable key rate and precluding real-time implementation.

We instead precompute these objects for a large, fixed set of configurations $\{\config_i\}$, which we call the database, while accepting a small mismatch $\Delta\config_i = \config_i - \config_\mathrm{exp}$. Each entry $\mathbf{v}_i$ depends only on the configuration $\config_i$ which is independent of the runtime session data $(\mathbf{c}, \lambdaEC(\mathbf{c},\mathbf{h}))$. Entries within a given operating point are therefore reusable across sessions, and no pre-session optimization is required. Evaluating \autoref{eq:VarLength} for a database entry on the session data $(\mathbf{c}, \lambdaEC(\mathbf{c},\mathbf{h}))$ then reduces to an inner product $\mathbf{f}_i\cdot\mathbf{c}$ plus scalar corrections, yielding a candidate key length $\ell_i$ at negligible runtime cost. On \autoref{fig:database} (b) and (c) we illustrate this approach.

A mismatch must not compromise security. The coefficient $\kappa_i$ has to remain a valid entropy bound for the true channel, so that $\ell_i$ lower-bounds the key length attainable at $\config_\mathrm{exp}$. Concretely, a larger assumed amplitude $\gamma_i \geq \gamma_\mathrm{exp}$ enlarges the signal available to Eve, making the optimized bound $\kappa_i$ more conservative. A smaller assumed electronic noise $\nu_{\mathrm{el},i} \leq \nu_{\mathrm{el,exp}}$ effectively increases the excess noise attributed to Eve's interference, again tightening the bound, while $\eta_{\mathrm{D},i} \geq \eta_{\mathrm{D,exp}}$ assumes more losses due to Eve's interference. Together, these ensure $\ell_i \leq \ell_\mathrm{exp}$  and the selected key length is always a valid lower bound; therefore, a mismatch costs key length, not security. We call a configuration satisfying this requirement valid. Validity can be certified directly from the calibrated parameters through a set of directional inequalities (see \autoref{app:database}), without performing the runtime optimization. The secret key length for a session is the maximum over valid entries,
\begin{equation} \label{eq:database_max}
    \ell := \max_{\{i \,:\, \config_i \text{ valid}\}} \ell_i .
\end{equation}
which is illustrated in \autoref{fig:database} (d).

The sensitivity of $\ell$ to a mismatch varies strongly across parameters, which determines how finely each axis of the grid is sampled. A detector-noise mismatch dominates and its axis is sampled finely, whereas the key length is far more robust to excess noise and channel distance, which are sampled coarsely. More details can be found in \autoref{app:database}.

\subsection*{Experimental implementation}

\begin{figure*}[h]
  \centering
  \includegraphics[width=\textwidth]{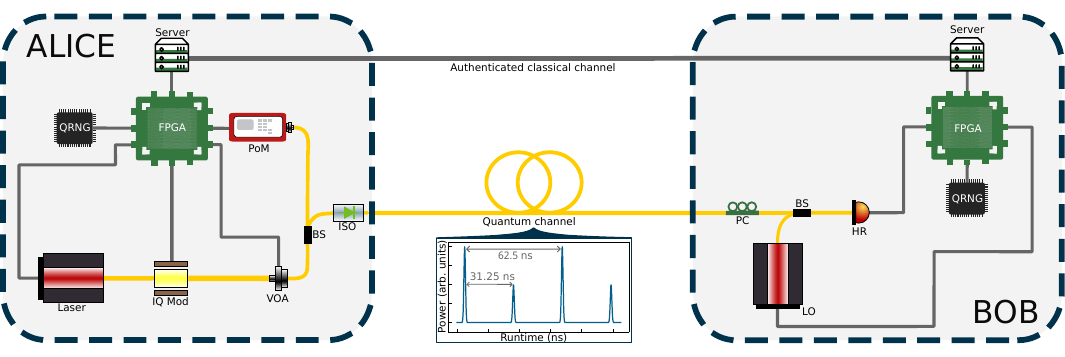}
  \caption{
    Experimental hardware overview. Dark links represent electrical cables, while yellow links represent optical fiber connections. The plot below the quantum channel showcases the presence of both reference and QPSK pulses, with its relative power being exaggerated for better visualization. Acronyms: VOA (Variable Optical Attenuator), IQ Mod (In-Phase and Quadrature component Modulator), BS (Beam Splitter), PoM (Power Meter), FPGA (Field-Programmable Gate Array), QRNG (Quantum Random Number Generator), PC (Polarization Controller), LO (Local Oscillator), HR (Heterodyne Receiver).}
  \label{fig:setup}
\end{figure*}

In Alice, a semiconductor laser emitting in the C-band in continuous operation is used to generate attenuated pulses by means of an IQ modulator. The amplitude of the reference pulses (see \hyperref[sec:methods]{Methods}) is set with a variable optical attenuator and is monitored with an optical power meter. Low-level digital signal processing (DSP) is handled by an FPGA, whereas a server controls the overall experiment, including the classical communication via the authenticated channel. The authenticated classical channel is established via Ethernet, and a standard optical fiber with 0.2 dB/km loss is used as the quantum channel. In Bob, another semiconductor laser is employed as a local oscillator to measure the pulses' amplitude and phase by means of a heterodyne receiver based on 3x3 optical coupling, as proposed in \cite{Adillon2025_3x3coupler, Sarmiento:26}. A polarization controller is employed to align the polarization of the received pulses with that of the local oscillator, in order to maximize the interference visibility. \autoref{fig:setup} provides an overview of the experimental setup, together with a plot showcasing the presence of both reference and QPSK pulses in the quantum channel.

\subsection*{Experimentally obtained secret key fractions}

\begin{figure*}
    \centering
    \includegraphics[width=\textwidth]{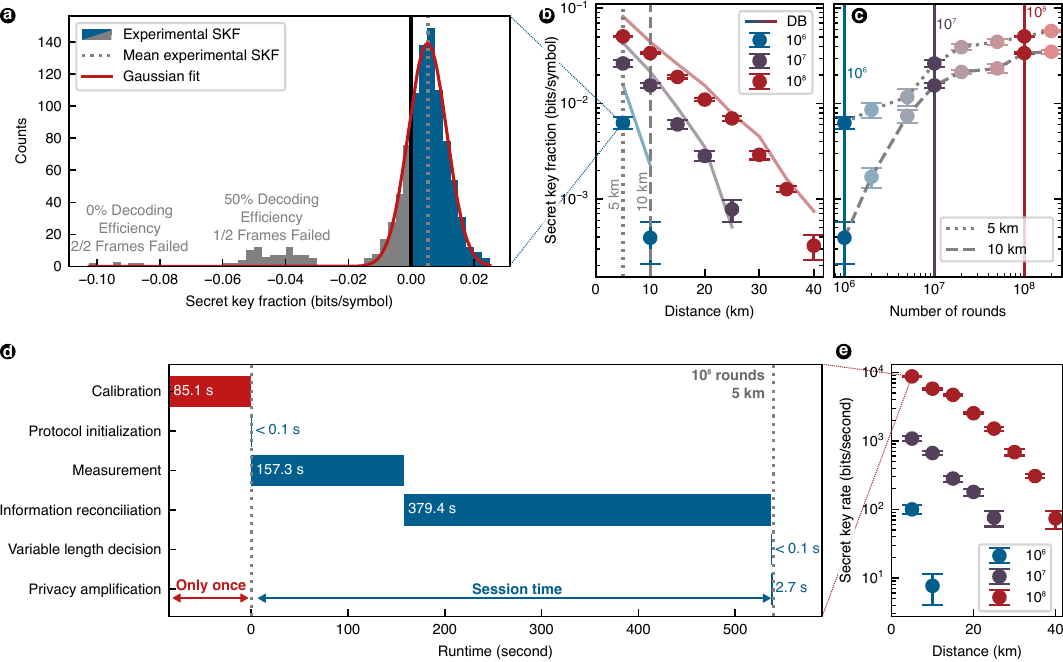}
    \caption{
        Experimental results.
        (a) Histogram of the experimentally achieved SKF distribution for \SI{5}{km} and \num{1e6} rounds. For calculating the mean, all negative values are set to zero before averaging. 
        (b) SKF as a function of distance for $n = \num{1e6}$, $\num{1e7}$, and $\num{1e8}$ rounds. Theoretical values (DB) correspond to the expectation value of the SKF for parameters listed in \autoref{tab:parameters_opt}, calculated for an assumed error correction efficiency of  \SI{97}{\percent}, a detection efficiency of $\etaD = \SI{55}{\percent}$, and an electronic noise level of $\nuel = \SI{77.5}{mSNU}$. As, on average, the parameters for the theoretical values are more favorable than those achieved experimentally, the plotted theoretical SKF typically sits above the experimental means. Error bars denote the standard error of the mean.
        (c) SKF as a function of the number of symbols for distances of \SI{5}{km} and \SI{10}{km}.
        (d) Runtime breakdown for a representative session at \SI{5}{km} and $n = \num{1e8}$ rounds. Calibration is performed once per system and amortized across subsequent sessions; the remaining stages constitute the per-session runtime. The measurement step includes shot and electronic noise calibration for Bob's detector.
        (e) Secret key rate in bits per second as a function of distance, obtained by dividing the secret key length by the per-session runtime.
    }
    \label{fig:results}
\end{figure*}
We applied the protocol over distances ranging from \SIrange{5}{40}{km} with $n$ between $10^6$ to ${2\times 10^8}$. \autoref{fig:results}\,(a) shows the secret key fraction (SKF) $\ell / n$ distribution at \SI{5}{km} for $n = 10^6$, where we use only two LDPC frames for error correction. The discrete decoding efficiency (\SI{0}{\percent}, \SI{50}{\percent}, or \SI{100}{\percent}, corresponding to 2, 1, or 0 failed frames) produces three distinct populations in the histogram. Only the \SI{100}{\percent} population yields positive key fractions and is well described by a Gaussian fit. Across the full distance range, the error correction efficiency $\beta_\mathrm{EC}$, defined as the ratio of the LDPC code rate to the Shannon limit, varies from \SI{86}{\percent} at short distances to \SI{96}{\percent} at long distances.

Each datapoint in \autoref{fig:results}\,(b) and (c) represents the mean SKF, with negative key fractions set to zero before averaging; error bars denote the standard error of the mean. The solid lines indicate expectation values for the SKF computed for a determined set of  parameters (see \autoref{tab:parameters_opt} in \autoref{app:database} for a complete list): an error-correction efficiency of $\beta_\mathrm{EC} = \SI{97}{\percent}$,
a detection efficiency of $\etaD = \SI{55}{\percent}$, and an electronic noise level of $\nuel = \SI{77.5}{mSNU}$. The parameters chosen for these curves are more favorable than the session-averaged experimental conditions, so the curves sit above the data on average rather than acting as a hard ceiling: individual sessions can exceed them when instantaneous detector/excess noise falls below the mean.

\subsection*{Secret key rate with full protocol runtime}

The SKFs reported above quantify the information-theoretic output per symbol, but the practical figure of merit for a deployed system is the secret key rate in bits per second, which folds in every processing step required to deliver a usable key. To assess our implementation under realistic operating conditions, we ran the complete protocol online and in sequence, with no offline post-processing. This contrasts with other CVQKD demonstrations, where computationally intensive signal processing is performed offline on stored data (see for instance Ref. \cite{Hajomer2025}).

\autoref{fig:results}\,(d) shows the runtime breakdown of a representative session at \SI{5}{km} with $n =10^8$ rounds. Calibration accounts for \SI{85.1}{s} but is performed only once and amortized across all subsequent sessions; for a system operating continuously, its contribution to the per-key cost is negligible. The per-session runtime is dominated by information reconciliation (\SI{379.4}{s}), followed by the measurement itself (\SI{157.3}{s}), while the protocol initialization, VLD, and privacy amplification together contribute less than \SI{3}{s}. The LDPC decoding therefore constitutes the principal bottleneck of the current implementation; reducing this cost (through dedicated hardware acceleration or more efficient codes) is the principal target for further improvement.

\autoref{fig:results}\,(e) shows the resulting secret key rate as a function of distance for the same sessions as in \autoref{fig:results}\,(b) and (c), obtained by dividing the secret key length by the per-session runtime. At \SI{5}{km} and $n = 10^8$ rounds, the system delivers approximately \SI{10}{kbit/s}, decreasing to approximately one hundred bits per second at \SI{40}{km}. Larger block sizes are favored because the SKF grows with $n$ as finite-size penalties shrink towards the asymptotic limit. The per-session runtime is close to linear in $n$ -- measurement and LDPC reconciliation both scale with the number of rounds -- but the fixed overheads (initialization, VLD, privacy amplification; together $< \SI{3}{s}$) amortize over larger blocks, so the per-key cost falls. The dominant driver of the gap between the $n=\num{1e6}$ and $n=\num{1e8}$ curves is the growth of the SKF.

\section*{Discussion}\label{sec:discussion}

With this work, we have achieved the first real-time experimental demonstration of a composable CVQKD protocol using coherent states under general security, finite-size effects, and detector imperfections. Our methodology avoids practical bottlenecks, such as secret key estimation, thanks to a variable-length model and a precomputed database of possible configurations.

The implementation allows continuous operation of the protocol with all steps running autonomously on the measurement hardware, without offline post-processing or user input. We observed the protocol to be robust against long-term fluctuations of experimental drifts and were able to run it uninterrupted for days, providing a promising outlook for potential commercial applications. This is a significant improvement in practicality compared to fixed-length approaches, which require sharply stable operating conditions to stay within the tolerance bounds set by their security framework.

Future work shall explore new paths of improvement, such as addressing the cost of authentication using keys from a previous QKD session \cite{ferradini2025definingsecurityquantumkey}, and incorporating `on-the-fly' announcements to allow the public exchange of classical information between Alice and Bob while also transmitting quantum signals \cite{arqand2025MEAT,tupkary2026rigorouscompletesecurityproof}. Another pressing improvement is lifting the so-called cutoff assumption \cite{Lin2019Asymptotic} via dimension reduction \cite{Upadhyaya2021Cutoff}, allowing the estimation of the secret key length without considering a bounded dimension for coherent states. Such a challenge could be solved by using the framework recently developed in Ref. \cite{NavarroPostsel2026}, which also considers postselection techniques that improve the secret key rate by sifting any noisy measurements by Bob, which may result in an advantage to Eve.

On the hardware side, we identify potential improvements on both the optical and electronic components. The impact of reducing the complexity of the optical components on performance should be investigated, which could also reduce the requirements for the DSP by the FPGA. The information reconciliation is currently the dominating runtime limitation. Future work may address further optimization with respect to efficiency, overall key rate performance, and costs.

\section*{Methods}\label{sec:methods}

\subsection*{Transmitter}\label{subsec:tx}
The transmitter generates QPSK-modulated coherent states by applying amplitude and phase modulation to the output of a free-running, amplitude-stabilized continuous-wave laser in the C-band. Pulses are produced at a repetition rate of 16\,MHz (temporal separation $\approx 62.5$\,ns). To support synchronization, weak quantum pulses are interleaved with higher-amplitude fixed-power pilot pulses assisting clock and phase recovery at the receiver. Low-level DSP is handled by an FPGA controlling modulator calibration and a quantum random number generator. High-level processing (error correction, privacy amplification, key-rate estimation, and classical communication) is performed by a server unit.

\subsection*{Receiver}\label{subsec:rx}
The receiver performs coherent detection by interfering the incoming quantum signal with a local oscillator from an amplitude-stabilized continuous-wave laser. A polarization controller maximizes interference visibility by aligning the polarization of each received signal with that of the local oscillator. Real-time frequency stabilization is maintained by monitoring the pilot-pulse phase. After heterodyne detection, the signal is digitized and processed by an FPGA-based high-speed DSP unit. Immediately after, a hard decision is applied to every received coherent state. To preserve security and efficiency, raw digitized analog-digital converter data are discarded immediately after categorization, leaving only categorical outcomes for subsequent processing. High-level DSP (error correction, privacy amplification, and classical communication) is performed by a server unit.

\subsection*{Communication channels}\label{subsec:channels}
Classical communication is established via standard Ethernet using transmission control protocols.
The quantum channel uses standard optical fiber; variable transmission distances are emulated using a variable optical attenuator.

As mentioned before, the authentication of the classical channel is a crucial step in the implementation of any QKD protocol. Besides enforcing that Eve cannot attack said channel beyond wiretapping, it also ensures that diverse steps of the protocol, such as the VLD, are symmetric (i.e. they can be performed by either Alice or Bob without affecting the final outcome).

This does not hold in case of an asymmetric authentication, where the honest parties may arrive at different results as they do not share the exact same public information. Recent works \cite{ferradini2025definingsecurityquantumkey,tupkary2026rigorouscompletesecurityproof} have provided a description of the security and performance of QKD under said context, as well as an analysis of the cost of authenticating the channel using the key from a previous QKD session. We leave such security analysis and implementation for future work, and consider only the case of perfect, symmetric authentication via a pre-shared key.

\subsection*{Calibration}\label{subsec:calibration}
Some protocol parameters ($\pkey$, $\Delta$, $\sdelta$, $\gamma$) can be chosen freely, and are typically optimized based on the expected channel performance ($\xi$, $\etaCh$). Because raw analog-digital converter data are discarded during categorization, these parameters cannot be inferred directly from stored quantum samples. Instead, we estimate the channel state from publicly disclosed symbols from preceding sessions. A machine-learning procedure determines the best-fit parameters by minimizing the mismatch between numerical simulation outputs and observed experimental statistics, treating excess noise and transmittance as free variables. The resulting estimates are used as operational assumptions for the subsequent QKD session.

\subsection*{Error correction}
Error correction is performed during the Information reconciliation step using low density parity check (LDPC) codes, described by means of sparse matrices which enable efficient error correction through iterative decoding. The specific LDPC code used in each session is chosen imposing a maximum reconciliation efficiency of 97\% over the Shannon limit since the performance drops above this threshold. Then the optimal code among a collection of LDPC codes with different code rates is chosen according to the measured signal-to-noise ratio on the quantum channel. 

Our decoding algorithm is based on the Sum-Product Decoding Algorithm introduced in \cite{LDPC_Rao} and using the estimated channel parameters as an initial guess of the Log-Likelihood Ratio of each symbol.

\backmatter

\bmhead{Acknowledgements}
We thank Jeison Tabares, Marco Cofano, Samael Sarmiento Hern\'andez, Elisabeth Llanos Pla and Pol Adillon for the fruitful discussions. This project has received funding from the European Union’s Digital Europe Programme under the projects QUARTER (101091588) and QUARTERNEXT (101305103), and from the European Innovation Council's Horizon Europe EIC Accelerator Programme under the project MIQRO (101161539), and the European Union (QSNP, 101114043). MN acknowledges funding from the Government of Spain (Severo Ochoa CEX2019-000910-S and FUNQIP), Fundació Cellex, Fundació Mir-Puig, Generalitat de Catalunya (CERCA program), the European Union’s Horizon Europe research and innovation programme under the MSCA Grant Agreement No. 101081441.

\bibliography{references}

\onecolumn

\appendix

\section{Secret key rate estimation}\label{Sec:SKREstimate}

In this appendix we provide a detailed description of the mathematical concepts required for the estimation of $f_{\config}$ and $\kappa_{\config}$, which eventually define a configuration $\mathbf{v}_{\config}$ and the secret key length according to \autoref{th:MainTheorem} in the main text. 

\subsection{Alice's marginal}
As a first step, we note that the prepare-and-measure scenario is fully equivalent to an entanglement-based approach thanks to the source-replacement scheme \cite{BBM92}. In the entanglement-based picture, Alice always prepares the same entangled state
\begin{align}
    \ket{\psi}_{AA'} = \sum_{x = \mathtt{00}}^{\mathtt{11}} \frac{1}{2} \ket{x}_A \otimes \ket{\gamma_x}_{A'},
\end{align}
where $\gamma_x \in \{e^{i 5 \pi/4}\gamma,e^{i 3\pi/4}\gamma,e^{i 7\pi/4}\gamma,e^{i \pi/4}\gamma\}$ for $\gamma \in \mathbb{R}$. She sends register $A'$ to Bob and applies a projective measurement on her register $A$ in order to steer the final state sent. This notation, in addition to simplifying the theoretical calculations, allows us to explicitly formulate Alice's marginal condition on the quantum state $\omega_{AB}$ shared by Alice and Bob 
\begin{align}
    \Tr_B[\omega_{AB}] &= \Tr_{A'}[\psi_{AA'}] = \frac{1}{4} \sum_{x,y = \mathtt{00}}^{\mathtt{11}} \ket{x}\bra{y} \bra{\gamma_y} \ket{\gamma_x}  := \sigma_A.
\end{align}
Namely, register $A$ is inaccessible to Eve since it never leaves Alice's laboratory, so it can be taken as a constraint on Eve's attack, which is a key ingredient to apply the MEAT.

\subsection{Information postprocessing}
In this section we formalize the description of the different registers held by Alice and Bob for one round $j \in \{1,...,n\}$. In the case of Alice, she applies a projective measurement on her register $A$. She records her measurement outcome in a register $X_j = x$. Equivalently, for the prepare-and-measure scenario, she draws two random bits $x \in \{\mathtt{00},\mathtt{11},\mathtt{01},\mathtt{10}\}$ that she stores as $X_j = x$, and sends the associated state.

On the other hand, Bob draws a bit $I_j$ with probabilities $(\pkey,1-\pkey)$ to decide which discretization to use after his heterodyne measurement. After obtaining a measurement outcome $y_j=|r_j|e^{i\theta_j}$,  he records the intermediate registers according to \autoref{fig:protocol} (b), such that

\begin{align}
    Y_j &= \begin{cases}
        \mathtt{00} & \text{if} \; \theta_j\in\left[\frac{3\pi}{2}, 2\pi \right) \land I_j=0\\
        \vdots \\
        \mathtt{11} & \text{if} \; \theta_j\in\left[ 0, \frac{\pi}{2}\right) \land I_j=0\\
        \mathtt{A} & \text{if} \; \theta_j\in\left[ 0, \frac{\pi}{2}\right) \land |r_j|\in \left[ 0,\sdelta\right) \land I_j=1\\
        \vdots \\
        \mathtt{D} & \text{if} \; \theta_j\in\left[ \frac{3\pi}{2}, 2\pi \right) \land |r_j|\in \left[ 0,\sdelta\right) \land I_j=1 \\
        \mathtt{E} & \text{if} \; |r_j|\in \left[ \sdelta,\Delta\right) \land I_j=1\\
        \mathtt{F} & \text{if} \; |r_j|\in \left[ \Delta,\infty\right) \land I_j=1\\
    \end{cases}
\end{align}
To ease the notation, let us consider that Bob splits this register into two. Respectively, for key generation and parameter estimation, we have
\begin{align}
    \zkg_j = \begin{cases}
        Y_j & \text{if} \; I_j = 0, \\
        \bot & \text{if} \; I_j = 1,
    \end{cases} && \zpe_j = \begin{cases}
        \bot & \text{if} \; I_j = 0 ,\\
        Y_j & \text{if} \; I_j = 1.
    \end{cases}
\end{align}
During parameter estimation, Bob reveals register $\zpe_j$ for every round, which also decides $I_j$ deterministically. Accordingly, Alice performs a split for key generation and parameter estimation
\begin{align}
    \xkg_j = \begin{cases}
        X_j & \text{if} \; I_j = 0, \\
        \bot & \text{if} \; I_j = 1,
    \end{cases} && \xpe_j = \begin{cases}
        \bot & \text{if} \; I_j = 0, \\
        X_j & \text{if} \; I_j = 1,
    \end{cases}
\end{align}
respectively. She always reveals $\xpe_j$. Provided the public announcements, Alice and Bob define $C_j =[ \xpe_j \zpe_j]$ as the set of all public information (with $I_j$ fully determined by $\zpe_j$, such that it can be omitted) related to round $j$, whose values are given by the alphabet $\mathcal{C}=\left \{\perp,\perp \right \} \bigcup(\left \{\mathtt{00},...,\mathtt{11}\right \}\times\left \{\mathtt{A},...,\mathtt{F} \right \})$. Similarly, let us define $\widetilde{\mathcal{C}}=\mathcal{C}\setminus\{(\perp,\perp)\}$ for genuine parameter-estimation rounds.

With the prior definitions, we can describe the quantum channel constituting a single round of the protocol from registers $AB$ to $\zkg C$ \cite{Navarro2025}.
\begin{align}\label{eq:Channel}
    \mathcal{M}_{AB\xrightarrow{} \zkg C}(\cdot) &= \pkey \sum_{z=\mathtt{00}}^{\mathtt{11}}\Tr [ \mathds{I}_A\otimes\widehat{R}_{B}^z(\cdot)] \ketbra{z}_{\zkg}\otimes\ketbra{\perp,\perp}_C \nonumber \\ 
    &\quad + \ (1- \pkey)\sum_{(x,z)\in\widetilde{\mathcal{C}}} \Tr [ \ket{x}\bra{x}_A\otimes R_{B}^z(\cdot)]\ket{\perp}\bra{\perp}_{\zkg} \otimes\ketbra{x,z}_C
\end{align}
where  $\{\widehat{R}_{B}^z\}_{z=\mathtt{00}}^{\mathtt{11}}$ and $\{R_{B}^z\}_{z=\mathtt{A}}^\mathtt{F}$ denote the POVMs for key generation and parameter estimation rounds, according to Bob's measurement for each type of round. These are explicitly given by the region operators \cite{Lin2019Asymptotic,PascualGarcia2025}, and the parameters $\Delta_\mathrm{S},\Delta$, together with the calibrated parameters $\eta_\mathrm{D}, \nu_\mathrm{el}$. We refer the reader to \cite{LinTrusted2020}  for their explicit formulation, including the trusted noise, according to the Fock basis.

\subsection{Numerical framework for $\kappa$}

Provided all the aforementioned tools, we can explicitly lower-bound $\kappa$ according to its definition as a ${H}^{\uparrow, f}_{\alpha}$-normalization constant \cite{kamin25MEATsecurity} 
\begin{align}
  \kappa
:= \inf_{\substack{\omega_{AB}\in\mathcal{D}( AB )\\
\text{s.t. }\Tr_{B}[\omega_{AB}]=\sigma_{A}}}
\; {H}_{\alpha}^{\uparrow,f} (\zkg | {C}E )_{\mathcal{M}(\omega)}.
\label{eq:kappa_inf}
\end{align}
Here, $\mc{D}(AB)$ denotes the set of all positive semidefinite matrices defined on registers $AB$, and we have the $f$-weighted Rényi entropy of order $\alpha$ which, for an arbitrary tradeoff function $f:\mathcal{C}\xrightarrow[]{}\mathbb{R}$ and $\alpha\in(0,1)\cup(1,\infty)$, we can expand the $f$-weighted Rényi entropy \cite[Definition 4.1]{arqand2025MEAT} as:
\begin{equation}
\tilde{H}^{\uparrow, f}_{\alpha}(Z^{KG}|CE)_{\mathcal{M}(\omega)}
:= \frac{\alpha}{1-\alpha}
\log \left(
\sum_{c \in \mc{C}} \mc{M}(\omega)_{|c} 2^{ \frac{1-\alpha}{\alpha} \left( -f(c) + \tilde{H}^{\uparrow}_{\alpha}(Z^{KG}|E)_{\mathcal{M}(\omega)|c} \right)}
\right),
\end{equation}
where $\mc{M}(\omega)_{|c}$ denotes the state after a classical conditioning on $C=c$ (see for instance \cite{PascualGarcia2025} for the definition), and we have a conditional sandwiched Rényi entropy on the right-hand side {given by definition as \cite{tomamichel2015quantum} 
\begin{equation}
    \tilde{H}^{\uparrow}_{\alpha}(A|E)_{\rho}:= \; \sup_{\sigma \in \mathcal{D}(E)} \; -{D}_\alpha\left( \rho_{AE} || \mathds{I}_{A}\otimes \sigma_E \right),
\end{equation}
with the sandwiched Rényi relative entropy
\begin{equation} \label{eq:QRenyiDiv}
    D_\alpha(\rho\|\sigma) =  \frac1{\alpha-1}\log{\frac{\Tr\left[(\sigma^\frac{1-\alpha}{2\alpha}\rho\sigma^\frac{1-\alpha}{2\alpha})^\alpha\right]}{\Tr[\rho]}}.
\end{equation}}
Using duality arguments, we can decompose the sandwiched Rényi entropy as in \cite[Appendix A]{Navarro2025}. Provided Eq. (101) from the same reference, we have

\begin{equation}
H^{\uparrow,f}_\alpha (Z^{KG}|CE)_{\mathcal{M}(\omega)}
    \geq \frac{\alpha}{1-\alpha} \log \left( \sum_{c \in \tilde{\mathcal{C}}} \mc{M}(\omega)_{|c} 2^{\frac{\alpha-1}{\alpha}f (c)} + p^\mathrm{K} 2^{\frac{\alpha-1}{\alpha} f ({\perp,\perp})} \Psi_\mu\left(\mathcal{G}(\omega),\mathcal{Z\circ G}(\omega) \right) \right) 
    \label{eq:tradeoff_renyi}
\end{equation}
where  $\mu=1/\alpha$ and $\Psi_\mu$ comes from reducing the formulation of the sandwiched conditional Rényi divergence, with the explicit form

\begin{align}
    \Psi_\mu(\rho,\sigma) &=\Tr\left[ \left( \sigma^{\frac{1-\mu}{2\mu}}\rho \sigma^{\frac{1-\mu}{2\mu}} \right) ^\mu \right]=\Big\|\sigma^\frac{1-\mu}{2\mu}\rho^\frac12\Big\|^{2\mu}_{2\mu}.
\end{align}
For said function, $\g(\omega)=K_G\omega K_G^\dagger$ denotes the coherent measurement and generation of secret key bits by Bob, where $K_G$ is the superoperator\footnote{We note that, actually, $\g$ and $\z$ act on an intermediate register which is eventually manipulated to form the actual key register $\zkg$. In order to avoid an excessive digression, we directly identify said intermediate register with $\zkg$ and defer the reader to \cite{chung2025,Navarro2025} for further information.} \cite{Lin2019Asymptotic,LinTrusted2020}
\begin{equation}
    K_G=\sum_{z = \mathtt{00}}^{\mathtt{11}}\ket{z}_{Z^{KG}}\otimes\mathds{I}_A\otimes\sqrt{\widehat{R}_B^z}.
\end{equation}
While $\z(\sigma)= \sum_{z = \mathtt{00}}^{\mathtt{11}} K_z\sigma K_z^\dagger$ constitutes a pinching map, which performs dephasing according to the superoperators
\begin{equation}
    K_z=\ketbra{z}_{\zkg} \otimes\mathds{I}_{AB}, \quad  z = \mathtt{00},...,\mathtt{11}.
\end{equation}
Next, we introduce a facial reduction \cite{drusvyatskiy2017,hu2022} which ensures that the final optimization problem is well-defined while reducing its numerical complexity. Following the process explained in  \cite[Section 4.1, Section 5.3]{Navarro2025}, we arrive at a simplified function 
\begin{equation}
    \widehat{\Psi}_\mu(\omega)=\left\| \z \circ \g (\omega)^{\frac{1-\mu}{2\mu}}K_G \omega^{\frac{1}{2}}\right\|_{2\mu}^{2\mu}.
\end{equation}
We may now substitute $\widehat{\Psi}_\mu$ in \eqref{eq:tradeoff_renyi} for $\Psi_\mu$. Then, replacing this on \eqref{eq:kappa_inf}, we obtain the expression

\begin{equation}
    \kappa\geq \inf_{\substack{\omega_{AB}\in\mathcal{D}( AB )\\
\text{s.t. }\Tr_{B}[\omega_{AB}]=\sigma_{A}}} \; \frac{\alpha}{1-\alpha} \log \left( \sum_{c \in \tilde{\mathcal{C}}} \mc{M}(\omega)_{|c} 2^{\frac{\alpha-1}{\alpha}f (c)} + p^\mathrm{K} 2^{\frac{\alpha-1}{\alpha} f (\perp,\perp)}\widehat{\Psi}_\mu(\omega)  \right) 
\label{eq:kappa_inf_2}
\end{equation}

\subsection{Conic formulation}
Making use of the conic formulation introduced in \cite{Navarro2025}, by defining the FastRényiQKD cone
\begin{equation}
    \mathcal{K}^\mu=\left\{ (u,\omega)\in\mathbb{R}\times\mathbb{H}_\succ^q; \; u\geq -\widehat{\Psi}_\mu(\omega)\right\},
\end{equation}
we can reformulate the inequality in Eq. \eqref{eq:kappa_inf_2} as the conic minimization problem
\begin{equation}\label{eq:Conic_kappaFast}
\begin{gathered}
    \kappa \geq  \min_{u, \omega} \frac{\alpha}{1-\alpha}  \log  \left(  \sum_{{c} \in \tilde{\mathcal{C}}} \mc{M}(\omega)_{|c} 2^{\frac{\alpha-1}{\alpha}f ({c})} - p^\mathrm{K} 2^{\frac{\alpha-1}{\alpha} f (\perp,\perp)} u \right) \\ %
        \mathrm{s.t.} \quad \Tr_B[\omega_{AB}] = \sigma_{A}, \\ 
    (u,\omega) \in \mathcal{K}^{\mu}.
\end{gathered}
\end{equation}
Thanks to the monotonicity of the negative logarithm, we can solve this optimization by simply maximizing the argument of the logarithm, which is affine and therefore readily solvable via non-symmetric conic programming (in particular, using the Skajaa-Ye algorithm \cite{skajaa2015,papp2017}).

Now, the only remaining step is how to find appropriate values for the tradeoff function $f$. Intuitively, the tradeoff function indicates the achievable scoring that Alice and Bob can assign to their measurements with respect to Eve in order to maximize the secret key. Although this function is arbitrary, it can be optimally chosen by adapting a fixed-length formulation \cite{kamin25MEATsecurity,tupkary2026rigorouscompletesecurityproof} into an honest implementation (where the finite frequency distribution coincides with the expected probability distribution).

This technique was introduced in \cite[Lemma 4.12]{arqand2025MEAT}, and for our formulation results in the conic program 

\begin{equation}\label{eq:Conic_2}
\begin{gathered}
    \min_{h_\mathrm{KL},h_\mathrm{QKD}, u, \lambda, \omega}  \frac{\alpha}{\alpha - 1} [h_\mathrm{KL} - p^\mathrm{K} h_\mathrm{QKD}] \\ %
    \mathrm{s.t.} \quad \Tr_B[\omega_{AB}] = \sigma_{A}, \\ 
    (u,\omega) \in \mathcal{K}^{\mu}, \\
    \left(h_\mathrm{KL}, \lambda, \mathcal{{M}}(\omega)_C \right) \in \mathcal{K}_\mathrm{KL}, \\
    \left(h_\mathrm{QKD},\Tr[\g(\omega)] ,- u \right) \in \mathcal{K}_{\log}, \\
    \sum_{c \in {\mc{C}}} \lambda(c) = 1,\; \lambda \ge 0, \\
    q - \lambda = 0.
\end{gathered}
\end{equation}
where we used the logarithmic and Kullback-Leibler cones, defined as
\begin{align}
    \mathcal{K}_\mathrm{KL} &= \left\{(h_{\mathrm{KL}},q,p) \in \mathbb{R} \times \mathbb{R}_>^d \times \mathbb{R}_>^d : h_{\mathrm{KL}} \geq \sum_{j=1}^d q(j) \log[q(j)/p(j)] \right\}, \\
    \mathcal{K}_{\log} &= \left\{ (h, v, u) \in \mathbb{R} \times \mathbb{R}_> \times \mathbb{R}_> : h \leq v \log\left(u/v\right) \right\}.
\end{align}
Solving the dual of the last constraint in \eqref{eq:Conic_2} provides the value for $f$.
In order to solve the minimizations related to $\kappa$ and $f$, we used the programming language Julia \cite{JuliaLang}, which allows conic optimization via the libraries Hypatia \cite{coey2022performance,coey2022solving} and JuMP \cite{Lubin2023}. In particular, Hypatia includes a standard definition of the logarithmic and Kullback-Leibler cones, together with the FastRényiQKD cone \cite{Navarro2025} provided through the ConicQKD package \cite{ConicQKD}, via an extension introduced in \cite{lorente2024}.

\section{Precomputed database construction} \label{app:database}

With the ideas presented above, we calculate one possible secret key length provided the statement of Theorem \ref{th:MainTheorem}. Now we illustrate how to calculate the set of all possible, effective configurations and build the database.

\subsection{Matrix formalism} \label{sec:matrix}
We recall \autoref{eq:VarLength}
\begin{align*}
        \ell_{\config} \leq \max\left\{0,\mathbf{v}_{\config} \cdot {\mathbf{w}}\right\}
\end{align*}
and replace the vector 
$\mathbf{v}_{\config}  \leftarrow  [\mathbf{v}_0, \dots, \mathbf{v}_m] := \mathbf{V}$ with a matrix $\mathbf{V}$ of $m+1$ valid configurations. As stated in the main text, all negative keys are set to zero, such that we arrive at the expression
\begin{align}
        \boldsymbol{\ell} \leq \mathbf{V} \cdot {\mathbf{w}} \enskip ,
\end{align}
with the $\cdot$ now representing a matrix-vector-multiplication and $\boldsymbol{\ell}$ a vector of length $m$; the inequality $\leq$ is guaranteed element-wise. We then  define
\begin{equation}
    \ell := \max \, \boldsymbol{\ell} \enskip ,
\end{equation}
which is equivalent to \autoref{eq:database_max} but computationally more efficient.

\subsection{Validity conditions}

Not all mismatches $\Delta \config_i = \config_i - \config_\mathrm{exp}$ between experimental configurations $\config_\mathrm{exp}$ and database configuration $\config_i$ lead to valid secret key length estimations $\ell_i$. We require that $\pkey$, $\sdelta$ and $\Delta$ must match exactly. Since $\kappa_i$ does not directly depend on the correctness parameter $\eEC$, the secrecy parameter $\ePA$, the number of rounds $n$, the excess noise $\xi$ and the channel transmittance $\etaCh$, we accept all mismatches for these parameters.

Alice's amplitude $\gamma$ enters $\kappa_i$ through her marginal state. As a general rule, we accept mismatches where database configurations $\config_i$ are more favorable from Eve's perspective than reality. For $\gamma$ this is the case if
\begin{subequations}
\begin{equation}
    \gamma_i \geq \gamma_\mathrm{exp} \enskip .
\end{equation}
The same general rule applies to the electronic noise $\nuel$ and Bob's detection efficiency $\etaD$, which affect $\kappa_i$ through Bob's POVM elements, and effectively control how much excess noise is allocated to Eve's interference. If the database is more favorable for Eve, i.e., if it would allocate more excess noise than reality, we consider the mismatch as valid. This is guaranteed if
\begin{align}
\nu_{\mathrm{el},i} &\leq \nu_\mathrm{el,exp}
\\
\eta_{\mathrm{D},i} &\geq \eta_\mathrm{D,exp} \enskip .
\end{align}
\end{subequations}

\subsection{Grid sampling and mismatch sensitivity}

\begin{figure*}
    \centering
    \includegraphics[width=\textwidth]{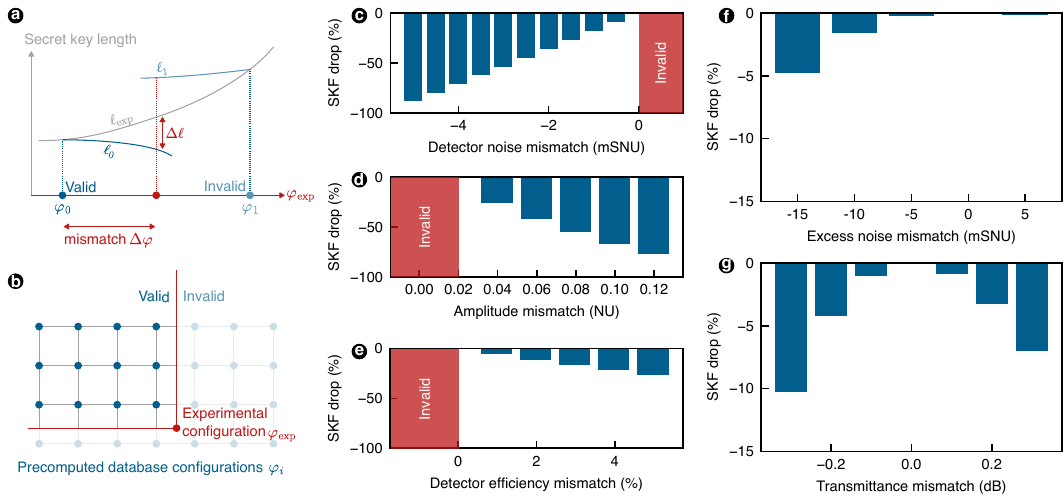}
    \caption{
        Mismatch sensitivity of the secret key length.
        (a) Evaluating the expected secret key length of a valid configuration $\config_0$ on the experimental statistics lower-bounds the secret key length obtained by evaluating the trade-off function directly at $\config_\mathrm{exp}$; the mismatch $\Delta\config = \config_i - \config_\mathrm{exp}$ reduces the key length by $\Delta\ell_i$. Invalid configurations $\config_1$ violate the lower-bound requirement.
        (b) A multi-dimensional grid of database configurations $\config_i$ (shown here in 2D for illustration) is precomputed. The experimental configuration $\config_\mathrm{exp}$ generally falls between grid points, and only a subset of the surrounding configurations are valid. Among these, the configuration yielding the highest secret key length is selected.
        (c)--(g) Simulated secret key length drop $\Delta\ell^{\mathrm{rel}}$ as a function of the mismatch in a single parameter, shown for $n = 10^6$ and \SI{5}{km} (channel transmittance \SI{1}{dB}) as a representative operating point; all other parameters are held to the values in \autoref{tab:parameters_const}. For example, (c) corresponds to $\Delta\config = [0, \dots, 0, \Delta\nuel]$. Validity restricts the admissible mismatch direction for $\nuel$ (c), $\gamma$ (d) and $\etaD$ (e), with invalid regions shaded; excess noise (f) and transmittance (g) are admissible in either direction.
    }
    \label{fig:sensitivity}
\end{figure*}

A valid mismatch, by design, assumes more favorable conditions for Eve and consequently will come at the cost of a reduced key length. In this section, we analyze this cost quantitatively in order to select the spacing of the pre-computed grid of configurations $\config_i$. We define the absolute (see \autoref{fig:sensitivity} a) and relative cost for a mismatch as
\begin{subequations}
\begin{align}
    \Delta\ell_i
    &:=
    \ell_i - \ell_\mathrm{exp}
    \\
    \Delta\ell^{\mathrm{rel}}_i
    &:=
    \frac{\Delta\ell_i}{\ell_{\mathrm{exp}}} \enskip .
\end{align}
\end{subequations}
\autoref{fig:sensitivity} (c) to (g) show the relative cost
as a function of the mismatch in a single parameter at a representative operating point ($\SI{5}{km}$, $n = 10^6$). The analysis should in general be repeated for every distance and block length used; the qualitative ranking of sensitivities is preserved across the explored range. A detector-noise mismatch (c) is the most consequential: a mismatch of only \SI{5}{mSNU} reduces the secret key length by nearly \SI{90}{\percent}, and this sensitivity grows with channel distance. The amplitude mismatch (d) exceeds \SI{50}{\percent} for severe values and the detection-efficiency mismatch (e) exceeds \SI{30}{\percent}, while excess noise and transmittance (f, g) are robust: mismatches of \SI{15}{mSNU} or \SI{0.3}{dB} cost less than \SI{10}{\percent} of the achievable key.

These observations inform the database sampling. The detector-noise axis is sampled finely, in \SI{0.5}{mSNU} steps. Because validity forces $\nuel \leq \nu_\mathrm{el,exp}$,
the worst-case half-step penalty is of order \SI{10}{\percent}, growing with distance, and is reduced further in practice as the selection picks the closest valid grid point. Excess noise and channel distance, to which the key length is far more robust, are sampled coarsely (\SI{5}{mSNU} and \SI{5}{km}) with negligible penalty. This non-uniform sampling reduces the database size relative to a uniformly fine grid, bringing the offline computation within practical reach. Computing the database entries required roughly two weeks of offline computation on a consumer-grade CPU (Intel Core i9-14900HX).

\subsection{Choice of controllable protocol parameters}

\begin{table}[h]
    \centering
    \caption{Chosen protocol parameters. The optimisations were performed using the constant parameters or their limits listed in \autoref{tab:parameters_const}. In addition, an excess noise level of $\xi \approx \SI{5}{mSNU}$ and a detector noise level of $\nuel \approx  \SI{77.5}{mSNU}$ was assumed.}
    \begin{tabular}{r r r r r r r r}
    \toprule
    Distance (km) & $n$ & $\gamma$ & $p_\mathrm{K}$ & $\Delta$ & $\sdelta$ & $\log_{10}(\alpha-1)$ \\
    \midrule
    0  & \num{1e6} & 0.86 & 0.58 & 5.0 & 1.7 & -2.99 \\
    5  & $''$      & 0.79 & 0.48 & 4.9 & 1.6 & -2.67 \\
    10 & $''$      & 0.75 & 0.30 & 4.3 & 1.5 & -2.37 \\
    \midrule
    5  & \num{1e7} & 0.86 & 0.74 & 5.9 & 1.6 & -3.70 \\
    10 & $''$      & 0.78 & 0.67 & 6.0 & 1.5 & -3.46 \\
    15 & $''$      & 0.77 & 0.60 & 6.6 & 1.6 & -3.35 \\
    20 & $''$      & 0.72 & 0.49 & 8.0 & 1.4 & -3.14 \\
    25 & $''$      & 0.70 & 0.30 & 6.9 & 1.4 & -2.90 \\
    30 & $''$      & 0.70 & 0.30 & 5.5 & 1.4 & -2.90 \\
    \midrule
    5  & \num{1e8} & 0.90 & 0.86 & 2.6 & 1.6 & -4.54 \\
    10 & $''$      & 0.84 & 0.84 & 2.6 & 1.5 & -4.40 \\
    15 & $''$      & 0.78 & 0.76 & 2.6 & 1.4 & -4.19 \\
    20 & $''$      & 0.78 & 0.76 & 2.6 & 1.4 & -4.10 \\
    25 & $''$      & 0.70 & 0.30 & 5.5 & 1.4 & -3.04 \\
    30 & $''$      & 0.73 & 0.50 & 5.5 & 1.6 & -3.72 \\
    35 & $''$      & 0.69 & 0.31 & 5.5 & 1.3 & -3.44 \\
    40 & $''$      & 0.68 & 0.30 & 5.5 & 1.3 & -3.47 \\
    45 & $''$      & 0.65 & 0.30 & 5.5 & 1.3 & -3.47 \\
    \bottomrule
    \end{tabular}
    \label{tab:parameters_opt}
\end{table}

When initially creating the database, some protocol parameters ($\gamma$, $\Delta$, $\sdelta$, $\pkey$) in the configuration $\config$ as well as $\alpha$ and $\mathbf{f}$ can be chosen freely. In practice, they will be optimized with respect to $\mathbb{E}[\ell]$ given the number of symbols $n$, the channel parameters ($\etaCh$, $\etaD$, $\xi$, $\nuel$), and the expected error correction performance $\beta_\mathrm{EC}$. \autoref{tab:parameters_opt} lists the results of this optimization performed for $n \in \{\num{1e6}, \num{1e7}, \num{1e8}\}$ and distances from \SIrange{0}{45}{km} in steps of \SI{5}{km}. The protocol parameters for missing configurations are interpolated using a nearest neighbor approach. In addition to protocol parameters, the values for the R\'enyi order $\alpha$ are also optimized simultaneously and listed in the same table.

\section{Information Reconciliation} \label{Sec:Leakage}
The information reconciliation is based on error correction methods based on low-density parity-check codes.

\subsection{Error correction leakage} 
Let $L_\mathrm{LDPC}$ be the code length and $2n \times \pkey_\mathrm{exp}$ be the total number of key bits. The latter will rarely be an exact multiple of $L_\mathrm{LDPC}$, so we write
\begin{equation} \label{eq:frames_remainder}
    2n \times \pkey_\mathrm{exp} = m \times L_\mathrm{LDPC} + \lambda_\mathrm{fit} \enskip ,
\end{equation}
where the divisor $m$ equals the number of LDPC frames and $\lambda_\mathrm{fit}$ is the remainder. As a conservative estimate, we treat all bits of the remainder $\lambda_\mathrm{fit}$ as fully leaked. We also define fit efficiency $\beta_\mathrm{fit} \in [0,1]$ via
\begin{equation} \label{eq:beta_fit}
    \beta_\mathrm{fit}
    :=
    \frac{m \times L_\mathrm{LDPC}}{2n \times \pkey_\mathrm{exp}}
    \enskip .
\end{equation}
which leads to
\begin{equation} \label{eq:leak_fit}
    \lambda_\mathrm{fit}
    \stackrel{\eqref{eq:beta_fit}}{=}
    (1 - \beta_\mathrm{fit}) \times 2n \times \pkey_\mathrm{exp}
    \enskip .
\end{equation}
Not all frames decode correctly. We define the decoding efficiency $\beta_\mathrm{dec} \in [0,1]$ as the fraction of the $m \times L_\mathrm{LDPC}$ fitted bits that belong to correctly decoded frames. The number of bits belonging to incorrectly decoded frames is then
\begin{equation} \label{eq:leak_dec}
    \lambda_\mathrm{dec}
    :=
    (1 - \beta_\mathrm{dec}) \times m \times L_\mathrm{LDPC}
    \stackrel{\eqref{eq:beta_fit}}{=}
    (\beta_\mathrm{fit} - \beta_\mathrm{dec} \times \beta_\mathrm{fit}) \times 2n \times \pkey_\mathrm{exp} \enskip ,
\end{equation}
all of which we again treat as fully leaked. The decoding success or failure of a frame is decided based on a per frame hash value of length $b_\mathrm{EC}$ bits exchanged between both parties. This adds another contribution 
\begin{equation} \label{eq:leak_hash}
    \lambda_\mathrm{hash}
    :=
    m \times b_\mathrm{EC}
\end{equation}
to the leakage. In particular, this cost ensures the correctness of the protocol, with a parameter $\varepsilon_\mathrm{EC}$ defining the condition 
\begin{align}
\varepsilon_\mathrm{EC} \geq m \times 2^{- b_\mathrm{EC}}.
\end{align}
Let now $R$ be the code rate of the LDPC code (in information bits per coded bit). The number of correctly decoded bits is given by $\beta_\mathrm{dec} \times m \times L_\mathrm{LDPC}$. The encoding process leaks
\begin{equation} \label{eq:leak_enc}
    \lambda_\mathrm{enc}
    :=
    (1 - R) \times \beta_\mathrm{dec} \times m \times L_\mathrm{LDPC}
    \stackrel{\eqref{eq:beta_fit}}{=}
    (\beta_\mathrm{dec} \times \beta_\mathrm{fit}  - R \times\beta_\mathrm{dec} \times \beta_\mathrm{fit})  \times 2n \times \pkey_\mathrm{exp} 
\end{equation}
bits of information. The total leakage is therefore defined as as the sum of all individual contributions
\begin{equation}
    \lambdaEC
    :=
    \lambda_\mathrm{enc} + \lambda_\mathrm{dec}+ \lambda_\mathrm{fit}+ \lambda_\mathrm{hash}
    \stackrel{\eqref{eq:leak_fit},\eqref{eq:leak_dec},\eqref{eq:leak_hash},
    \eqref{eq:leak_enc}}{=}   
    (1 - R \times \beta_\mathrm{dec} \times \beta_\mathrm{fit}) \times 2n \times \pkey_\mathrm{exp} + m \times b_\mathrm{EC}.
\end{equation}

\subsection{Shannon limit}

With Gray encoding, the uniform QPSK protocol factorises into two identical and independent binary symmetric channels, one per encoded bit. We account for leakage per channel use, i.e.\ per encoded bit; the total of $2n \times \pkey_\mathrm{exp}$ key bits then corresponds to $2n \times \pkey_\mathrm{exp}$ channel uses. Denoting $\hat{Z}^\mathrm{KG}$ as Bob's key register without the symbol $\perp$, the conditional entropy of a single such channel, in bits per channel use, is
\begin{equation} \label{eq:cond_entropy}
    H(\hat{Z}^\mathrm{KG}\!\mid \!\xkg) = h_2(e) \enskip ,
\end{equation}
where $h_2(e) = -e\log_2 e - (1-e)\log_2(1-e)$ is the binary entropy function and $e$ the bit-error rate. The capacity\footnote{Note that this capacity corresponds to the hard-decision binary-symmetric-channel; $f_\mathrm{EC}$ and $\beta_\mathrm{EC}$ are therefore benchmarked against the hard-decision capacity rather than the soft-information capacity of the underlying channel.} per channel use is the mutual information
\begin{equation}
  C = I(\xkg;\hat{Z}^\mathrm{KG}) = 1 - H(\hat{Z}^\mathrm{KG}\!\mid\!\xkg) = 1 - h_2(e) \enskip .
\end{equation}
Defining the encoding efficiency $\beta_\mathrm{enc} := R/C$ (with both $R$ and $C$ now expressed per channel use, so that $\beta_\mathrm{enc} \in [0,1]$ and $\beta_\mathrm{enc} = 1$ at the Shannon limit) and the overall error-correction efficiency
\begin{equation}
    \beta_\mathrm{EC} := \beta_\mathrm{enc} \times \beta_\mathrm{dec} \times \beta_\mathrm{fit} \enskip,
\end{equation}
we obtain, using $\beta_\mathrm{EC} \, C = R \, \beta_\mathrm{dec} \, \beta_\mathrm{fit}$,
\begin{equation}\label{eq:leak_shannon}
  \lambdaEC = (1 - \beta_\mathrm{EC} \times C) \times 2n \times \pkey_\mathrm{exp} + m \times b_\mathrm{EC} \enskip .
\end{equation}

\subsection{Conversion from error correction efficiency to scaling parameter}
An alternative, widely used parametrisation of the error-correction cost expresses the leakage
\begin{equation}\label{eq:leak_f}
    \lambdaEC
    =
    f_\mathrm{EC} \times 2n \times \pkey_\mathrm{exp} \times H(\hat{Z}^\mathrm{KG}\!\mid\!\xkg)
    \stackrel{\eqref{eq:cond_entropy}}{=}
    f_\mathrm{EC} \times 2n \times \pkey_\mathrm{exp} \times h_2(e) \enskip ,
\end{equation}
via a scaling parameter $f_\mathrm{EC} \geq 1$ with respect to the conditional entropy. Here, we consider the per-frame hash term $m \times b_\mathrm{EC}$ is negligible against the leading contribution in the regimes of interest and is dropped in the conversion below. Equating \eqref{eq:leak_shannon} (without the hash term) and \eqref{eq:leak_f} yields the conversion
\begin{subequations}
\begin{align}
  f_\mathrm{EC}
  &=
  \frac{1 - \beta_\mathrm{EC}\bigl(1 - h_2(e)\bigr)}{h_2(e)} \enskip,
  \\
  \beta_\mathrm{EC} &= \frac{1 - f_\mathrm{EC}\,h_2(e)}{1 - h_2(e)} \enskip.
\end{align}
\end{subequations}


\end{document}